\documentclass[twocolumn,showpacs,showkeys,preprintnumbers,amsmath,amssymb]{revtex4}

\usepackage{xcolor}
\usepackage[active]{srcltx}
\usepackage{graphicx}
\usepackage{dcolumn}
\usepackage{bm}
\usepackage{epsfig}
\usepackage{epstopdf}
\usepackage{physics}

\begin{document}

\preprint{IST/UL.2022-M J Pinheiro}

\title[From Phase Space to Non-Equilibrium Dynamics: Exploring Liouville's Theorem and its Implications]{From Phase Space to Non-Equilibrium Dynamics: Exploring Liouville's Theorem and its Implications}

\author{Mario J. Pinheiro}
\email{mpinheiro@tecnico.ulisboa.pt}

\affiliation{Department of Physics, Instituto Superior T\'{e}cnico - IST, Universidade de Lisboa - UL, Av. Rovisco Pais, \&
1049-001 Lisboa, Portugal}

\homepage{https://www.researchgate.net/profile/Mario-Pinheiro}

\thanks{}

\date{\today}

\begin{abstract}
The Liouville theorem is a fundamental concept in understanding the properties of systems that adhere to Hamilton's equations. However, the traditional notion of the theorem may not always apply. Specifically, when the entropy gradient in phase space fails to reach equilibrium, the phase-space density may not have a zero time derivative, i.e., $\frac{d\rho}{dt}$ may not be zero. This leads to the concept of the set of attainable states of a system forming a compressible "fluid" in phase space. This observation provides additional insights into Hamiltonian dynamics and suggests further examination in the fields of statistical physics and fluid dynamics. In fact, this finding sheds light on the limitations of the Liouville theorem and has practical applications in fields such as beam stacking, stochastic cooling, and Rabi oscillations, among others.
\end{abstract}



\maketitle

\section{Introduction}

A key finding in classical mechanics, Liouville's theorem~\cite{Liouville1838} has significant effects on how classical systems behave. The volume of phase space covered by a system experiencing Hamiltonian motion is preserved throughout time, according to the theorem. If we think about the six-dimensional phase space that is covered by the positions and velocities of every particle in a classical system, the volume of this phase space that the system occupies will not change over time. This has several important implications: (i) Trajectories in phase space cannot intersect, according to the conservation of phase space volume. As a result, there are no "random" changes in the state of the system; rather, the evolution of a system in phase space is entirely predictable~\cite{hp1890am,poincare1892}; (ii) According to Liouville's theorem, a system may develop from any initial state to any final state over the course of time in phase space~\cite{Arnold,Arnold1989,Marsden1994,Gallavotti}. This is caused by the fact that phase space volume is conserved, which protects the system from "losing" information as it develops; (iii) The behavior of macroscopic systems is significantly impacted by the conservation of phase space volume. For instance, it suggests that when a gas of non-interacting particles evolves, the particle distribution in phase space stays constant~\cite{Toda}. The classical Boltzmann distribution, which defines the statistical behavior of classical systems at equilibrium, is based on this principle~\cite{Ehrenfest}.

Liouville's theorem is an effective tool for understanding how classical systems behave in phase space. It forms the basis for classical mechanics' deterministic development and contributes to the explanation of the statistical behavior of macroscopic systems. We can find a description of Liouville's theorem close to the original in the well-known book of Paul Ehrenfest~\cite{Ehrenfest}.

The Schr\"{o}dinger quest for a quantum mechanical description based on the eikonal form of the Maxwell set of equations dictated their type of solutions based on amplitudes, the superposition principle for their solutions, and the typical interferences effects that are the outcome of squaring amplitudes. In a result of this path, the mathematical framework is founded in linear algebra and Hilbert spaces. The fitting of quantum mechanical theory to classical mechanics was made by forcing the relationship between the Poisson brackets and
the commutator brackets, $[A,B]=i \hbar \{A,B\}$. If instead, the starting point was classical mechanics, then a different outcome would result, mainly no amplitude to intensity relationships and the superposition principle would be applied to forces (or potentials).

Liouville's theorem provides a crucial link between classical and quantum mechanics through the concept of phase-space density. In classical mechanics, phase space is a six-dimensional space spanned by the position and momentum coordinates of all particles in the system. The phase-space density function, denoted by $\rho(q,p,t)$, gives the probability density of finding a system in a particular phase-space point $(q,p)$ at time $t$.

In quantum mechanics, the phase space is replaced by the Hilbert space of the system, and the concept of the phase-space density is replaced by the wave function, denoted by $\Psi(q,t)$ and the probability density of finding a quantum system in a particular state is given by $|\Psi(q,t)|^2$.

The Wigner function is a mathematical tool that provides a link between the classical and quantum descriptions of the phase-space density, defined as:

\begin{multline}
W(q,p,t) = \frac{1}{2\pi\hbar}\int_{-\infty}^{\infty}\Psi^*\left(q-\frac{\lambda}{2},t\right)\Psi\left(q+\frac{\lambda}{2},t\right) \\ e^{-i\frac{p\lambda}{\hbar}}d\lambda.
\end{multline}

The Wigner function is a real-valued function that satisfies a version of Liouville's theorem known as the Wigner-Liouville equation, $ \frac{\partial W}{\partial t} + \{H,W\} = 0 $, where $H$ is the Hamiltonian of the system and $\{\cdot,\cdot\}$ denotes the Poisson bracket. The Wigner-Liouville equation describes the time evolution of the Wigner function and provides a bridge between the classical and quantum descriptions of the system.

Furthermore, Liouville's theorem provides the foundation for the correspondence principle, which states that the behavior of quantum systems should reduce to classical behavior in the limit of large quantum numbers. This is achieved through the use of coherent states, which are quantum states that exhibit classical-like behavior. The Wigner function for coherent states approaches the classical phase-space density in the limit of large quantum numbers.

In quantum mechanics, the non-conservation of phase space implies that the uncertainty principle is fundamental and cannot be overcome. This is because, in quantum mechanics, the phase-space density is related to the probability amplitude of the system, which determines the probability of finding the system in a particular state. The non-conservation of phase space means that the probability density of the system cannot be conserved, which implies that the uncertainty principle must hold. The uncertainty principle states that the more precisely the position of a particle is known, the less precisely its momentum can be known, and vice versa. Therefore, the non-conservation of phase space implies that the uncertainty principle is an essential feature of quantum mechanics and cannot be eliminated.

There have been many experiments that have confirmed the uncertainty principle, and it is considered to be a fundamental principle of nature~\cite{Aspect,Hensen}. There have been attempts to find ways around the uncertainty principle, such as using entangled particles or non-local measurements, but these approaches have not been successful in violating the principle~\cite{Cunha,Kimble,Aspect1982}. The uncertainty principle is deeply embedded in the foundations of quantum mechanics and is a necessary consequence of the wave-particle duality of quantum objects.

Hence, the nonconservation of phase space has significant implications for classical and quantum mechanics. Classical mechanics challenges the Liouville theorem and suggests the compressibility of phase space. In quantum mechanics, it implies the uncertainty principle, demonstrating limits to measurement precision, and has important implications for interpretation and experiment design.

The aim of this work is to demonstrate that the assumption that Hamiltonian systems obey the Liouville theorem may be challenged by the inclusion of Eqs. (8)-(9). This is due to the fact that if entropy gradients in the phase-space do not equilibrate, the time derivative of the phase-space density may not be zero. Consequently, the set of states that a system could access forms a volume in phase space denoted by $\Gamma$, which is analogous to a compressible fluid.

\section{The equation of motion for physical quantities}

The equation of motion for any arbitrary physical quantity is
\begin{equation}\label{eq1}
\dot{F}=\frac{\partial F}{\partial q}\dot{q}+\frac{\partial F}{\partial p}\dot{p}=[H,F],
\end{equation}
with the hamiltonian $H\equiv p \dot{q}-L(\dot{q},q)$, $[,]$ representing the Poisson brackets, and the Hamiltonian equations of motion
\begin{eqnarray}\label{eq2}
\dot{q}=\frac{\partial H}{\partial p} &;& \dot{p}=-\frac{\partial H}{\partial q}.
\end{eqnarray}

The fundamental equation of motion in Hamiltonian mechanics is given by Eq. \eqref{eq1}, which provides a mathematical description of the time evolution of any physical quantity. This equation expresses the rate of change of a physical quantity $F$ as a function of the generalized coordinates $q$ and momenta $p$. The Hamiltonian $H$, which is defined in terms of the Lagrangian $L$ as $H\equiv p \dot{q}-L(\dot{q},q)$, plays a central role in the formulation of Hamiltonian mechanics, and is related to the total energy of the system. The Poisson brackets $[,]$, defined as $[A,B]\equiv \sum_i(\frac{\partial A}{\partial q_i}\frac{\partial B}{\partial p_i}-\frac{\partial A}{\partial p_i}\frac{\partial B}{\partial q_i})$, capture the algebraic structure of Hamiltonian mechanics and provide a powerful tool for the calculation of physical quantities.

The Hamiltonian equations of motion, given by Eq. \eqref{eq2}, are a set of first-order differential equations that describe the time evolution of the generalized coordinates and momenta. These equations are derived from the Hamiltonian $H$ and express the principle of least action in Hamiltonian form. They relate the time derivatives of the generalized coordinates and momenta to the partial derivatives of the Hamiltonian with respect to the conjugate variables.

Hamiltonian mechanics provides a powerful framework for the description of classical and quantum systems and has a wide range of applications in physics and engineering. The formalism is particularly useful for the analysis of conservative systems, where the total energy is conserved.

\section{Out-of-equilibrium dynamics}

A new form of canonical momentum and equation of motion has been derived from a recent variational principle~\cite{Pinheiro}. The fundamental equation of motion has the form of a local balance equation with the spatial gradient of entropy as the source term. The gradient of the total entropy in momentum space is also given, which, when maximized, leads to the total canonical momentum. The new set of equations of motion resembles the Hamiltonian formulation of dynamics and complies with the Helmholtz free energy. Liouville's equation has been defined, along with the Liouville operator, and the expectation value of a function has also been given.

In that framework, the new form of canonical momentum is:
\begin{equation}\label{eq3}
\frac{\partial \bar{S}}{\partial \mathbf{p}^{(\alpha)}} \geq 0
\end{equation}
and the fundamental equation of motion:
\begin{equation}\label{eq4}
\frac{\partial \bar{S}}{\partial \mathbf{r}^{(\alpha)}} = -\frac{1}{T} \pmb{\nabla}_{r^{(\alpha)}} U^{(\alpha)} - \frac{1}{T} m^{(\alpha)} \frac{\partial \mathbf{v}^{(\alpha)}}{\partial t} \geq 0.
\end{equation}

Eq.~\ref{eq4} gives the fundamental equation of dynamics and has the form of a general local balance equation having as source term the spatial gradient of entropy, $\nabla_a S>0$, whilst Eq.~\ref{eq3} gives the canonical momentum. At thermodynamic equilibrium, the total entropy of the body has a maximum value. In the more general case of a non-equilibrium process, the entropic gradient must be positive in both Eqs.~\ref{eq3}-~\ref{eq4}. The interplay between energy-minimizing tendencies and entropy maximization may introduce new physics through a set of two first-order differential equations. These equations have the potential to reveal novel insights into the underlying dynamics of physical systems.

In non-equilibrium processes, the gradient of the total entropy in momentum space multiplied by factor $T$ is given by
\begin{equation}\label{6}
\frac{\partial \overline{S}}{\partial \mathbf{p}^{(\alpha)}} = \frac{1}{T} \left\{
-\frac{\mathbf{p}^{\alpha)}}{m^{(\alpha)}} + \frac{q^{(\alpha)}}{m^{(\alpha)}}\mathbf{A} +
\mathbf{v}_e + [\mathbf{\omega} \times \mathbf{r}^{(\alpha)}] \right\},
\end{equation}
so that maximizing entropy change in Eq.~\ref{eq3} leads to the well-known total canonical momentum:
\begin{equation}\label{eq7}
\mathbf{p}^{(\alpha)} = m^{(\alpha)} \mathbf{v}_e + m^{(\alpha)} [\mathbf{\omega} \times
\mathbf{r}^{(\alpha)}] + q^{(\alpha)} \mathbf{A}.
\end{equation}

The above formulation bears some resemblance with the Hamiltonian formulation of dynamics which expresses first-order constraints of the Hamiltonian $H$ in a $2n$ dimensional phase space, $\dot{\mathbf{p}}=-\partial H/\partial \mathbf{q}$ and  $\dot{\mathbf{q}}=\partial H/\partial \mathbf{p}$, and can be solved along trajectories as quasistatic processes, revealing the same formal symplectic structure shared by classical mechanics and thermodynamics. The sharing of a formal symplectic structure between classical mechanics and thermodynamics implies a common geometric framework for the equations of motion, which enables the application of Hamiltonian mechanics to the study of thermodynamic systems and suggests the existence of underlying mathematical structures that are common to many different physical systems.

In the context of our approach, the new set of equations of motion should read:
\begin{eqnarray}
  \mathbf{\dot{p}} &=& -\pmb{\nabla} H + T \pmb{\nabla} \overline{S} =-\frac{\partial}{\partial \mathbf{q}} (H-T\overline{S})\\ \label{eqp1}
  \mathbf{\dot{q}} &=&  -T\pmb{\nabla} \overline{S} +\pmb{\nabla} H= \;\;\frac{\partial}{\partial \mathbf{p}} (H-T\overline{S}). \label{eqp2}
\end{eqnarray}
We have identified $U$ as equivalent to $H$, and it is worth noting that the motion of the system is now governed by the Helmholtz free energy, $\mathfrak{H} = H-T\overline{S}$, rather than just the Hamiltonian alone.
The gradients of the system's Hamiltonian function and the thermodynamic quantities are connected to the time derivatives of the system's position and momentum by the equations. There is an easy way to connect the macroscopic thermodynamic parameters of temperature, entropy, and energy to the microscopic characteristics of the system's particles due to the identification of the Hamiltonian function with the Helmholtz free energy, H = U - TS. The reformulation may be a useful tactic for statistical mechanics' study of the behavior of complex systems, with several applicability in physics, chemistry, and materials science, as we will suggest later.

Our investigation centers around Liouville's equation
\begin{equation}\label{eq8}
\frac{d \rho}{dt}=\imath [\rho,H],
\end{equation}
where the function $\rho(q,p,t)$ is defined in a way such as the product
\begin{equation}
    \rho(q,p,t) dq dp = \rho(q,p,t) d \Omega
\end{equation}
represents the number of system points in the phase volume $d\Omega$ around the point 
$(q,p)$ at the time $t$.
We can write
\begin{equation}\label{eq9}
\imath \frac{\partial \rho}{\partial t}=L \rho
\end{equation}
where
\begin{equation}\label{eq10}
L=-\imath \frac{\partial H}{\partial p}\frac{\partial }{\partial q}+\imath \frac{\partial H}{\partial q}\frac{\partial}{\partial p},
\end{equation}
represents the Liouville operator (and $\imath = \sqrt{-1}$).
\begin{equation}\label{eq11}
\langle A \rangle =\int dp dq A(p,q) \rho
\end{equation}
\begin{equation}\label{eq12}
\frac{\partial \rho}{\partial t}=-\partial_p H \partial_q \rho + \partial_q H \partial_p \rho
\end{equation}
But from Eqs. we have
\begin{equation}\label{eq13}
\frac{\partial \rho}{\partial t}=-\dot{q}\partial_q \rho - \dot{p}\partial_p \rho -T\partial_p \overline{S}\partial_q \rho + T \partial_q \overline{S} \partial_p \rho
\end{equation}
or
\begin{equation}\label{eq14}
\frac{d \rho}{dt}=-T \left[ \partial_p \overline{S} \partial_q \rho -\partial_q \overline{S} \partial_p \rho \right].
\end{equation}
If we introduce now the usual Poisson bracket for two variables $A$ and $B$:
\begin{equation}\label{eq15a}
    [A,B]=\sum_i \left(\frac{\partial A}{\partial q_i}\frac{\partial B}{\partial p_i}-\frac{\partial B}{\partial p_i}\frac{\partial A}{\partial q_i} \right),
\end{equation}
we could express the Liouville equation in a more comprehensive form.
\begin{equation}\label{eq15}
\frac{\partial \rho}{\partial t}+\pmb{u}\cdot \pmb{\nabla} \rho = -T[\partial_p \overline{S}\partial_q  - T \partial_q \overline{S}\partial_p] \rho.
\end{equation}
Note that
\begin{equation}\label{eq16}
\pmb{u}\cdot \pmb{\nabla}=\sum_l \left(\frac{\partial H}{\partial p_l}\frac{\partial }{\partial q_l} - \frac{\partial H}{\partial q_l}\frac{\partial}{\partial p_l} \right).
\end{equation}
Eq.~\ref{eq16} implies a correction to the Liouville equation, that needs to be written now under the form:
\begin{equation}\label{eq17}
\frac{\partial \rho}{\partial t}=[H,\rho]-T[\overline{S},\rho],
\end{equation}
or
\begin{equation}\label{eq18}
\frac{d \rho}{dt}=-T[\overline{S},\rho] \neq 0.
\end{equation}
But, as the free energy can be defined by $\overline{F}=F_0-T\overline{S}$, we may drop out the approximation of isothermic states and write instead a more general form of Liouville's theorem for out-of-equilibrium systems:
\begin{equation}
    \frac{d \rho}{dt}=-[\overline{F},\rho] \neq 0.
\end{equation}
Non-isothermic states involves a delicate balance between energy and entropy, which can make their analysis difficult and may not be necessary for understanding the behavior of out-of-equilibrium systems. By redefining the free energy as $\overline{F}=F_0-T\overline{S}$, we can instead focus on a more general form of Liouville's theorem that applies to a broader range of physical systems.

The introduction of Equations (7)-(8) necessitates a reformulation of Liouville's theorem. When the entropy gradients in phase space fail to equilibrate, the time derivative of the phase-space density, $d\rho/dt$, may not be zero, resulting in a set of possible system states forming a compressible "fluid" volume in phase space denoted as $\Gamma$. This observation sheds light on why the Liouville theorem does not seem to hold in certain techniques, such as beam stacking, electron cooling, stochastic cooling, synchrotron radiation, and charge exchange~\cite{Ruggiero,Winston_2009}, or in the Boltzmann equation when the collision operator is irreversible~\cite{Villani}. Ref.~\cite{Datsko_2016} reports that Liouville's theorem is insufficient for predicting the behavior of the atmosphere and climate, even in the case of a simple linear oscillator.

This failure of Liouville's theorem has significant technological implications, and we provide examples of such cases below.

\subsection{Relaxation of an initially out-of-equilibrium system towards thermal equilibrium}

Given the previously defined operators, the Helmholtz operator for a system can be expressed as:

$$\hat{H} - T\hat{S} = \hat{U} - TS(\hat{\rho}) = (\hat{\mathcal{H}} - \mu \hat{\mathcal{N}}) - T\hat{S}(\hat{\rho})$$

Here, $\hat{\mathcal{H}}$ is the Hamiltonian operator, $\mu$ is the chemical potential operator, $\hat{\mathcal{N}}$ is the particle number operator, and $\hat{S}(\hat{\rho})$ is the operator for the entropy of the system. Note that the entropy operator is a function of the density operator $\hat{\rho}$, which describes the statistical distribution of the system. the von Neumann entropy of a density operator $\rho$ can be defined as $S(\rho) = -\mathrm{Tr}(\rho \ln \rho)$, where $\mathrm{Tr}$ denotes the trace. This is a standard expression for entropy in the context of quantum mechanics. 

To obtain a quantum field equation from the set of classical equations of motion given in Eqs. (7) and (8), we need to promote the variables p and q to quantum operators and replace the classical Poisson brackets with quantum commutators. We also need to introduce a time-dependent parameter $\lambda$ to control the transition from the classical to the quantum regime, such that $\lambda = 0$ corresponds to the classical limit, and $\lambda = 1$ corresponds to the fully quantum regime. This can be done using the so-called Wigner-Weyl transformation, which maps classical variables to quantum operators.

Let us define the Wigner function as:
\begin{equation}
W(q, p, t) = \int dy e^{i p y / \hbar} \psi(q - y / 2, t) \psi^{*}(q + y / 2, t),
\end{equation}
where $\psi(q, t)$ is the wave function of the system. The Wigner function is a quasi-probability distribution (it may contain negative values) that encodes both the position and momentum information of the system and it satisfies the following properties: i) $W(q, p, t)$ is real; ii) $W(q, p, t)$ is normalized: $\int dq dp W(q, p, t) = 1$.

The marginal distributions of W(q, p, t) with respect to q and p recover the probability density and current of the system, respectively:
\begin{equation} \label{eq1}
\begin{split}
\rho(q, t) & = \int dp W(q, p, t) \\
 J(q, t) & = \int dp \frac{p}{m} W(q, p, t)
\end{split}
\end{equation}
where $m$ is the mass of the system. Using the Wigner function, we can rewrite the classical equations of motion in Eqs. (7) and (8) as:
\begin{equation}
\frac{\partial W}{\partial t} = \{H - TS, W\}
\end{equation}
where $\{A, B\}$ denotes the Poisson bracket of $A$ and $B$, and we have replaced the classical variables p and q with their corresponding quantum operators, $\hat{p}$ and $\hat{q}$. The Poisson bracket can be replaced with a commutator in the limit of large quantum numbers, $\hbar \to 0$, using the correspondence rule, $\{A, B\} \to  (1/\hbar) [\hat{A}, \hat{B}]$, where $[\hat{A}, \hat{B}]$ denotes the commutator of $\hat{A}$ and $\hat{B}$. So, the quantum field equation is:
\begin{equation}\label{eq26}
\frac{\partial \hat{W}}{\partial t} = [\hat{H} - T\hat{S}, \hat{W}],
\end{equation}
where $\hat{H}$, $\hat{S}$, and $\hat{W}$ are the quantum operators corresponding to the classical functions $H$, $S$, and $W$, respectively.

Eq.~\ref{eq26} is the quantum field equation in the Wigner representation. It describes the time evolution of the Wigner function of the system and can be used to calculate various properties of the system, such as its energy spectrum ( distribution of energy levels), correlation functions (relationship between different properties), and coherence properties (the degree to which the phases of different parts of a wave or system are related to one another).

Note that this definition of entropy assumes that $\rho$ is a positive semi-definite operator with a trace equal to one, which is the case for density operators. To define the entropy for a general operator, we need to consider more general definitions. The von Neumann entropy of a density operator $\rho$ is defined as $S(\rho) = -Tr(\rho \ln \rho)$. On the other hand, the Wigner function is a quasi-probability distribution function that represents the quantum state of a system in phase space, defined as:
\begin{equation}
W(q, p) = \frac{1}{\pi \hbar} \int \psi^{*}(q+y/2) \psi(q-y/2) e^{(-ipy/\hbar)} dy,
\end{equation}
where $\psi$ is the wave function of the system, $q$ and $p$ are the position and momentum variables. To apply the von Neumann entropy to the Wigner function, we first need to convert the Wigner function into a density operator (represented by a Hermitian matrix, and containing information about the probabilities of different outcomes of a measurement of the system). This can be done using the Wigner-Weyl transform:
\begin{equation}
 \rho = \frac{1}{\pi \hbar} \int W(q,p) D(q,p) dq dp
\end{equation}
where $D(q,p)$ is the Weyl operator, defined as:
\begin{equation}
 D(q,p) = e^{(i(p \wedge q - q \wedge p)/\hbar)}.
\end{equation}
Once we have the density operator $\rho$, we can calculate its von Neumann entropy using the formula $S(\rho) = -Tr(\rho \ln \rho)$, where $\ln \rho$ is the matrix logarithm of $\rho$, which can be obtained by diagonalizing $\rho$ and taking the logarithm of its eigenvalues. The trace $Tr(\rho \ln \rho)$ can then be evaluated by summing the diagonal elements of $\rho \ln \rho$. Hence, we can apply the von Neumann entropy to the Wigner function by first converting it into a density operator using the Wigner-Weyl transform, and then using the formula $S(\rho) = -Tr(\rho \ln \rho)$ to calculate its von Neumann entropy.

To apply the equation $d \rho/dt = -[F, \rho]$ to study the evolution of the phase space density during a system's transition, we must first define the free energy function F as a function of parameters that describe the transition. Suppose the transition is controlled by a parameter $\lambda$. In this case, the free energy function can be expressed as:
\begin{equation}
    F(\lambda) = H - \lambda G,
\end{equation}
where $H$ and $G$ are Hermitian operators that represent the Hamiltonian and some other observable, respectively. As $\lambda$ is varied, the system undergoes a transition from one phase to another, and we want to study the evolution of the phase space density as this happens. We can start by writing the Liouville equation in terms of the free energy function $F(\lambda)$:
\begin{equation}
 \frac{d \rho}{dt} = -[F(\lambda), \rho].
\end{equation}
Expanding the commutator, we get:
\begin{equation}
\frac{d \rho}{dt} = -FH \rho + H \rho F + \lambda GH \rho - \lambda H \rho G   
\end{equation}
Now, we can write the density operator $\rho$ as a sum of its eigenstates:
\begin{equation}
\rho  = \sum_n p_n |n><n|,
\end{equation}
where $p_n$ is the probability of finding the system in the nth eigenstate $|n>$. By writing the density operator as a sum of its eigenstates, we can then determine the probabilities of finding the system in each of its eigenstates at a given time.

Substituting this into the Liouville equation and using the orthonormality of the eigenstates, we get:

\begin{multline}
\frac{d}{dt}(p_n |n><n|) = -FH(p_n |n><n|) + \\ H(p_n |n><n|)F + 
\lambda GH(p_n |n><n|) - \\ \lambda H(p_n |n><n|)G,
\end{multline}

which simplifies to:
\begin{equation}
 \frac{d}{dt}(p_n) = -p_n (F_nn - F_ii) + \lambda p_n (G_nn - G_ii)   
\end{equation}
where $F_{nn}$ and $F_{ii}$ are matrix elements of the Hamiltonian in the nth and ith eigenstates, respectively, and similarly for $G_{nn}$ and $G_{ii}$.

The set of equations presented describes how the probabilities of the different eigenstates of the system, represented by $p_n$, evolve over time as the parameter $\lambda$ is varied. By solving these equations, we can analyze the behavior of the system as it undergoes the transition. One application of these equations is to calculate the average value of an observable $A$ in the different phases. This can be done using the formula:
\begin{equation}
  <A> = \sum_nn p_n <n|A|n>  
\end{equation}
where the sum is over the eigenstates of the system, and $<n|A|n>$ is the expectation value of the observable $A$ in the nth eigenstate.

\begin{figure}
    \centering
    \includegraphics[width=0.5\textwidth]{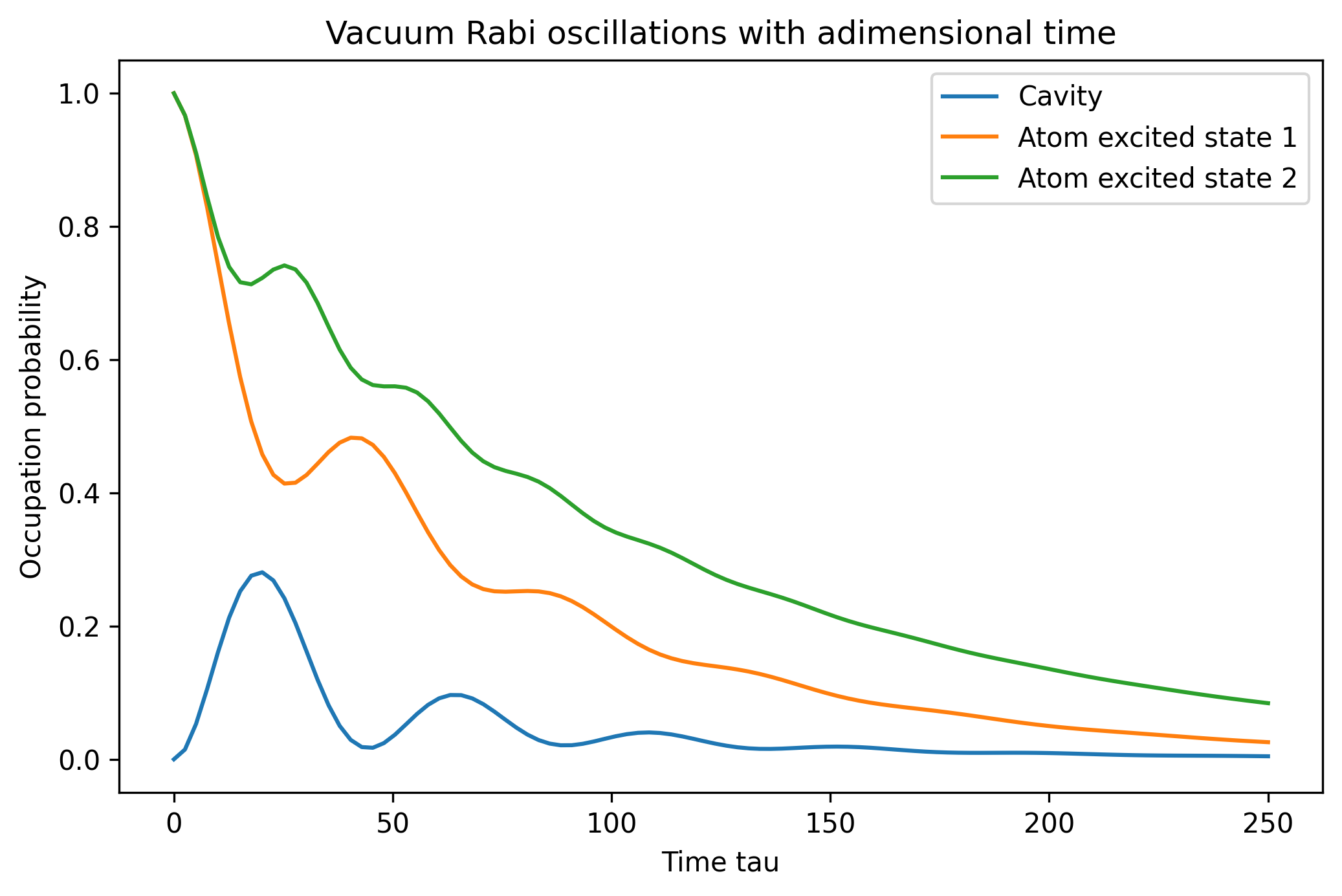}
    \caption{Visual representation of the interdependence of energy and entropy in Vacuum Rabi oscillations. Orange line: $\omega_1=0.90$, green line: $\omega_2=1.20$, blue line: $\omega_c = 1.0$ is the cavity frequency.}
    \label{Fig2}
\end{figure}

The inclusion of temperature and entropy in the Hamiltonian can lead to effects such as decoherence and relaxation, which can significantly influence the Rabi oscillations in a quantum system. When a quantum system interacts with a thermal environment, energy exchange, and dephasing result in a reduced coherence of the Rabi oscillations. Fig.~\ref{Fig2} illustrates one application of the previously described model, after scaling the energies of the cavity $\omega_c=1$ and atoms frequencies $\omega_1, \omega_2$~\cite{pinheiro2023rabi}. In a variety of quantum computing systems, these oscillations are essential for implementing quantum gates, which are the fundamental operations needed to build quantum circuits and execute quantum algorithms. 

We now have a theoretical framework, relying on the equation $d\rho/dt = -[F, \rho]$, which permits us to explore the time-dependent evolution of the phase space density as a system undergoes a transition. This framework is based on the analysis of the probabilities associated with the different eigenstates of the system.

\subsection{Nonelastic collisions between particles}
According to Eq.~\ref{eq18} the invariance of volume for canonically conjugated variables is not verified which implies that, in the presence of entropy gradients or out-of-equilibrium systems, there is no conservation of momentum nor kinetic energy (see also Ref.~\cite{Leontovitch}) in particle collisions.

This means that in the presence of entropy gradients or out-of-equilibrium systems, there is no conservation of momentum or kinetic energy in particle collisions. This lack of conservation of these quantities is significant because it implies that the usual assumptions made in equilibrium systems, where entropy gradients are absent, do not hold in out-of-equilibrium systems. This result is not new and has been discussed previously in the literature. Reference [5] is also cited as a source of further information on the subject. The lack of conservation of momentum and kinetic energy in particle collisions in out-of-equilibrium systems has important implications for understanding the behavior of systems far from equilibrium, such as those found in many biological, ecological, and social systems. Understanding and modeling such systems require a more complex and nuanced approach than is necessary for equilibrium systems, where the assumption of entropy gradients is usually valid.

\subsection{Brightness of an atomic beam Source}

Subjecting the axial or transverse velocity components of the beam to dissipative cooling dramatically compresses the phase space of the atom flux, resulting in dense, well-collimated atomic beams suitable for the study of atom optics, atom holography, or ultracold collision dynamics. Prodan et al.~\cite{Prodan} first demonstrated the importance of this phase-space compression. In fact, atomic beams can now achieve a level of “brightness” (atom beam flux
density per unit solid angle) many times greater than the phase-space conservation limit
imposed by the Liouville theorem (cf. Pierce~\cite{Pierce}, Sheehy et al.~\cite{Sheehy}, Kuyatt~\cite{Kuyatt}). The importance of dissipative cooling in compressing the phase space of atomic beams leads to dense, well-collimated atomic beams that are useful for studying various fields of physics, such as atom optics, atom holography, and ultracold collision dynamics. The phase-space compression was first demonstrated by Prodan et al.~\cite{Prodan} in 1994, highlighting its importance in the field of atomic physics.

Moreover, recent advancements in atomic beam technology have resulted in achieving "brightness" levels (atom beam flux density per unit solid angle) that surpass the phase-space conservation limit imposed by the Liouville theorem, which describes the conservation of phase-space volume in a classical dynamical system. This breakthrough is significant because it opens up new opportunities for studying the behavior of atomic beams in various applications, including materials science, quantum optics, and precision measurements. The references cited in the text provide further information on the research related to this topic.

\subsection{The mechanics of magnetic helicity in the plasma}

The helicity associated with ions and electrons in plasma has opposite signs. This is because helicity is a measure of the handedness of the magnetic field, and the ions and electrons have opposite charges and therefore move in opposite directions in a magnetic field. As a result, the magnetic fields generated by the two populations will have opposite handedness. In the expression for the free energy that includes both ions and electrons, we need to take into account the opposite signs of the helicity densities. A more general expression for the free energy that accounts for this is:
\begin{equation}
 F = \frac{B^2}{2 \mu_0} + (\alpha_i - \alpha_e) K,
\end{equation}
where $\alpha_i$ and $\alpha_e$ are the helicity densities associated with the ion and electron populations, respectively, and $K$ is the kinetic energy density. The term $(\alpha_i - \alpha_e)$ accounts for the opposite signs of the ion and electron helicities.

Taking the gradient of $F$ with respect to position, we obtain:
\begin{equation}
\grad F = \frac{1}{2 \mu_0} \grad (B^2) + (\alpha_i - \alpha_e) \grad (K).
\end{equation}
Using the same identity as before, $\grad (B^2) = 4 \alpha \mathbf{B}$, where $\alpha$ is the total helicity density, including contributions from both ions and electrons, we can write:
\begin{equation}
\grad F = \left(\frac{4 \alpha}{\mu_0} \right) B^2 + (\alpha_i - \alpha_e)  \grad (K).
\end{equation}
The total helicity density in plasma physics is a measure of the twistedness or knotting of magnetic field lines, and it's a conserved quantity in ideal magnetohydrodynamics (MHD). The gradient of free energy is related to the total helicity density, but with an additional term that accounts for the difference between ion and electron helicities. This term reflects how the dynamics of ion and electron populations can impact the overall plasma behavior.

Mathematically, the total helicity density is defined as the volume integral of the dot product between the magnetic field $\mathbf{B}$ and its vector potential $\mathbf{A}$, i.e., $\alpha = \int_V (\mathbf{A} \cdot \mathbf{B})dV$, where $V$ is the volume of the plasma. In general, the total helicity density can be both positive and negative, depending on the orientation and topology of the magnetic field lines.

\begin{figure}
    \centering
    \includegraphics[width=0.5\textwidth]{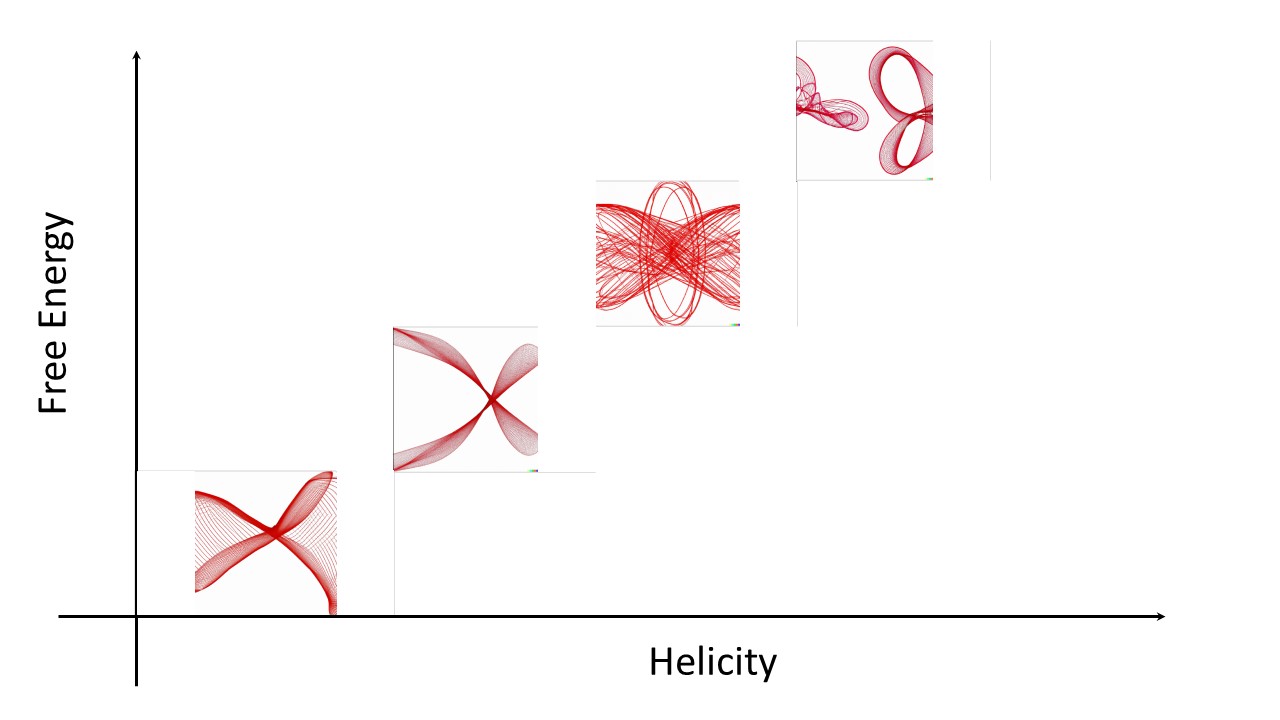}
    \caption{The relationship between free energy and helicity in magnetic fields, visualized through the increasing complexity of magnetic field lines as their writhe and twist increases. As the helicity of the field increases, so does its free energy, leading to more complex and tangled field lines.}
    \label{Fig1}
\end{figure}

In plasma, the total helicity density is related to the free energy of the system and plays a crucial role in determining the stability and dynamics of the plasma. The gradient of the total helicity density is related to the Lorentz force that acts on the plasma, and it can drive various instabilities and reconnection events in the magnetic field. Therefore, the total helicity density is an important quantity in plasma physics and is often used in theoretical and experimental studies of plasmas. In a plasma, both ions and electrons can contribute to the helicity of the magnetic field. The ion and electron helicities are defined as the volume integrals of the dot products between the magnetic field and the velocity of the respective species, i.e., $\alpha_i = \int_V (\mathbf{v}_i \cdot \mathbf{B}) dV$ and $\alpha_e = \int_V (\mathbf{v}_e \cdot \mathbf{B}) dV$, where $\mathbf{v}_i$ and $\mathbf{v}_e$ are the velocities of the ions and electrons, respectively. The total helicity density is the sum of the ion and electron helicities, i.e., $\alpha = \alpha_i + \alpha_e$. As shown in Figure \ref{Fig1}, the complexity of magnetic field lines increases with their writhe (measure the total amount of coiling or twisting in a knot) and twist (a measure of the local twisting or rotation of a knot).

Therefore, the total helicity density includes contributions from both ions and electrons and reflects the overall twistedness or knotting of the magnetic field lines in the plasma.

The difference between the ion and electron helicities, i.e., $(\alpha_i - \alpha_e)$, is related to the dynamics of the ion and electron populations in the plasma. If the ion and electron populations have different velocities or distributions, they can contribute differently to the helicity of the magnetic field and create a net helicity difference. This net helicity difference can in turn affect the stability and dynamics of the plasma and can lead to various instabilities or reconnection events in the magnetic field, as shown in~\cite{Myers2022} with the effect of energy conversion and dynamics of magnetic reconnection. Therefore, the ion and electron helicities, as well as their difference, are important quantities in plasma physics and can provide valuable insights into the behavior and evolution of plasmas.

The equation $d \rho/dt = -[F, \rho]$ describes the evolution of helicity density $(\rho)$ in plasma, where changes in free energy $F$ are linked to changes in magnetic field line twist and writhe. Increases in free energy can lead to increases in helicity density and vice versa. The equation provides a fundamental link between free energy and magnetic topology and highlights the important role of free energy in determining magnetic dynamics in plasma.

To illustrate the relationship between free energy, helicity, twist, and writhe, let us consider a simple example of a magnetic field in a plasma that has both twist and writhe. We can write the magnetic field in terms of its vector potential, A, as $\mathbf{B} = \grad \times \mathbf{A}$, and the helicity density of this magnetic field can be written as $\rho = \mathbf{A} \cdot \grad \times \grad{A}$. 
This equation relates the helicity density to the vector potential and its curl, and it quantifies the amount of twisting and linking of the magnetic field lines in the plasma.

The free energy of plasma, defined as the energy stored in the magnetic field, can be expressed as $F = \frac{1}{2 \mu_0} \int B^2 dV$. Perturbations in the magnetic field can affect the twist and writhe of the magnetic field lines and cause changes in the helicity density and free energy of the plasma. Adding twist to the magnetic field increases the helicity density and, in turn, the free energy of the plasma. The time derivative of the helicity density can express this relationship.
\begin{equation}
\frac{d \rho}{dt} = \int (\grad \times \mathbf{A}) \cdot \left( \grad \times \frac{d \mathbf{A}}{dt} \right) dV
\end{equation}
Using the equation of motion for the plasma, which is given by:
\begin{equation}
 \rho \mathbf{v} = [\mathbf{j} \times \mathbf{B}] - \epsilon_0 \frac{d \mathbf{E}}{dt},
\end{equation}
where $v$ is the plasma velocity, $j$ is the current density, $E$ is the electric field, and $\epsilon_0$ is the electric permittivity of free space, we can rewrite the time derivative of the helicity density as:
\begin{equation}
\frac{d \rho}{dt} = -2  \int (\mathbf{j} \cdot \mathbf{B}) dV.
\end{equation}
This equation shows that the time rate of change of the helicity density is proportional to the current density and the magnetic field. Thus, if we increase the twist in the magnetic field, we will also increase the current density, which will in turn increase the rate of change of the helicity density, and hence the free energy of the plasma.
Similarly, if we perturb the magnetic field by adding a small amount of writhe to it, the helicity density will again increase, and this will lead to an increase in the free energy of the plasma. This can be seen by considering the writhe of the magnetic field lines, which is given by:
\begin{equation}
Wr = \int \left[\mathbf{B} \cdot \grad \times \left(\frac{\mathbf{B}}{B^2} \right) \right] dV,
\end{equation}
where the integral is taken over the volume of the plasma. This equation quantifies the degree of linking of the magnetic field lines, and it is related to the helicity density through the equation $\rho = 2  Wr$, the greater the degree of linking between magnetic field lines, the higher the helicity density in the system.

Thus, we can see that changes in the free energy of the plasma can lead to changes in the twist and writhe of the magnetic field lines and that the helicity density provides a fundamental link between these quantities.

\section{Conclusion}

We conclude that, if the Liouville theorem reflects the properties of systems obeying Hamilton's equations, in our approach, introducing Eqs.(7)-(8) this is not necessarily so. If the gradients of entropy in phase space do not equilibrate, then $d\rho/dt$ is not necessarily null, which means that the set of states that a system can possibly attain form a volume in the phase-space $\Gamma$ representing a ``fluid" that may be compressible. The conclusion drawn is that the introduction of Eqs.(7)-(8) may invalidate the assumption that the Liouville theorem reflects the properties of systems obeying Hamilton's equations. The reason for this is that, if the gradients of entropy in phase space do not equilibrate, then the time derivative of the phase-space density may not necessarily be zero. This, in turn, means that the set of states that a system can possibly attain form a volume in phase space $\Gamma$ representing a "fluid" that may be compressible.

This result may be significant because it challenges the conventional understanding of the behavior of systems obeying Hamilton's equations and implies that the dynamics of such systems may be more complex than previously thought. Moreover, the idea that the phase space of a system may be compressible has important implications for understanding the thermodynamics of such systems and may have applications in fields such as statistical physics and fluid dynamics (due to non-conservative forces, such as turbulence or viscous dissipation).

\section*{Competing Financial Interests}
No targeted financial assistance was provided for this research by any public, private, or non-profit entity.
\bibliographystyle{unsrt}

\bibliographystyle{apsrev}
\end{document}